\documentclass[journal]{IEEEtran}

\usepackage{cite}      

\usepackage{graphicx}  

\begin{document}

\title{Phase-sensitive interrogation of fiber Bragg grating resonators for sensing applications}

\author{Jong H. Chow,
        Ian C.M. Littler\dag,
        Glenn de Vine,
        David E. McClelland,
        and~Malcolm~B.~Gray
\thanks{This research was partially supported by the Australian Research Council (ARC) under the auspices of the Australian Consortium for Interferometric Gravitational Astronomy, and partially funded with the assistance of the ARC Centres of Excellence Program and the ARC Federation Fellowship Scheme through the Centre for Ultrahigh Bandwidth Systems (CUDOS).  CUDOS is an ARC Centre of Excellence.}%
\thanks{The authors are with the Centre for Gravitational Physics, Faculty of Science, The Australian National University, Canberra, ACT 0200, Australia; \dag Ian C.M. Littler is with CUDOS (Centre for Ultrahigh Bandwidth Devices for Optical Systems), School of Physics, A28, University of Sydney, Camperdown, NSW 2006, Australia.}}

\markboth{Journal of Lightwave Technology,~Vol.~, No.~,~}{Shell \MakeLowercase{\textit{Jong H. Chow et al.}}: Phase-sensitive interrogation of fiber Bragg grating resonators for sensing applications}

\maketitle

\begin{abstract}
We discuss a phase-sensitive technique for remote interrogation of passive Bragg grating Fabry-Perot resonators.  It is based on Pound-Drever-Hall laser frequency locking, using radio-frequency phase modulation sidebands to derive an error signal from the complex optical response, near resonance, of a Fabry-Perot interferometer.  We examine how modulation frequency and resonance bandwidth affect this error signal.  Experimental results are presented that demonstrate when the laser is locked, this method detects differential phase shifts in the optical carrier relative to its sidebands, due to minute fiber optical path displacements.
\end{abstract}


\IEEEpeerreviewmaketitle

\section{Introduction}
 
\PARstart{T}{here} has been a large body of work involving fiber Bragg grating sensors over the past two decades \cite{FBGsensorsSurvey,FBGsensorsKersey,Othenos,Allsop}.  Early demonstrations were based on changes in the gross Bragg wavelength as the gratings were perturbed due to strain and temperature.  As interrogation techniques became more sophisticated, various signal processing and active fringe side locking schemes were employed, which dramatically improved their resolution \cite{Allsop,Dandridge,Song,Lissak,Arie,Fisher}.  This was further enhanced by refinement of grating design, enabling Fabry-Perot resonators to be fabricated \cite{Canning,LeBlanc,LoadLeBlanc,Rao,Wan}, which effectively multiply the phase change due to fiber optical path displacements.  With careful control of the grating writing process and appropriate choice of glass material, a Bragg grating fiber Fabry-Perot (FFP) can now have a finesse of well over 1000 and a linewidth of a few MHz \cite{Gupta}.  A fiber distributed feedback (DFB) laser can be fabricated when the FFP is written in a fiber amplifier and pumped optically \cite{Kringlebotn}.  These lasers have attracted significant interest for use in various schemes \cite{Lovseth,Frank} as active sensing elements, where changes in lasing wavelength due to environmental perturbations are used as the sensor signal.

The past decade has also seen intense international effort in attaining direct gravitational wave detection, which demands unprecedented interferometric sensitivity to measure the strain of space-time.  Towards achieving this ultra resolution, the Pound-Drever-Hall (PDH) laser frequency locking scheme \cite{Drever,Slagmolen,Black} is widely used.  It is adopted for laser frequency stabilization \cite{Day,Bondu}, interferometer longitudinal control, as well as gravitational wave signal extraction \cite{Strain,Muller,Mason,Shaddock}.

\begin{figure}
\centering
\includegraphics[width=3.0in]{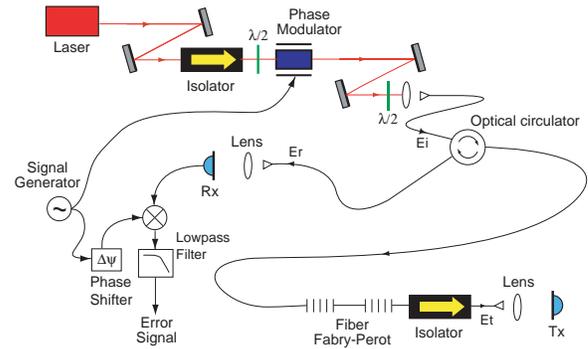}
\caption{Topology of our PDH fiber Fabry-Perot interrogation experiment.  Ei, Er and Et are the input, reflected and transmitted fields of the fiber Fabry-Perot, respectively.  Rx and Tx are photodetectors for reflected and transmitted light, and half-wave plates are denoted with $\lambda/2$.}
\label{schematic}
\end{figure}

While the PDH frequency locking technique is well-established with free-space bulk-optical resonators and solid-state lasers within the gravitational wave community, it can readily be extended to diode laser stabilization \cite{Schoof}, and guided-wave optics.  It has previously been utilized in a fiber laser stabilization scheme, where an Erbium doped fiber laser was referenced to a coated micro resonator \cite{Park}.  The PDH locking scheme can be adapted for both low frequency ($<$10Hz) quasi-static strain sensing, and dynamic measurements at higher frequencies.

In this paper we will discuss this technique in some detail, and demonstrate PDH locking for signal extraction in a fiber sensor, with a Bragg grating FFP as the sensing resonator.  Remote interrogation of passive resonators has significant advantages over active DFB lasers as sensing devices.  The problems relating to dopant clustering and relaxation oscillations in Erbium doped DFB lasers \cite{Delevaque,Sanchez,Loh1,Ding1,Ding2,Amroun} are completely avoided, and the undesirable effects of optical pump noise is eliminated \cite{Man}.  In addition, the interrogating laser can be housed in a controlled environment, and any residual introduction of phase noise due to laser intensity fluctuations are secondary effects.  When the chosen interrogating wavelength is 1550nm, telecoms grade SMF-28 fiber can be used for both laser delivery as well as grating sensor fabrication, with the dual benefit of low cost and low loss.  This removes the need for more exotic fibers which requires cutoff below the pump wavelength for single-mode pump delivery.  At 0.1-1mW output power, Erbium doped fiber DFB lasers are inherently inefficient compared with commercially available extra-cavity diode lasers, used in this work, with $>$10mW of output power.  This higher laser power improves the signal to shot noise ratio by up to an order of magnitude.  While this output power can potentially be matched by Er/Yb codoped DFB lasers, they require complex fiber geometry to achieve sufficient photosensitivity \cite{Dong,Carter}.  In addition, frequency instability due to thermal pump absorption continues to limit their sensing performance \cite{Man}.  Remote interrogation, therefore, presents itself as an elegant and superior sensing solution.

\section{Fiber Fabry-Perot complex response, resonance bandwidth, phase modulation frequency, and error signals}
For the purpose of this discussion, we will simplify our treatment of the FFP as similar to that of a free space resonant cavity, ie, within the bandwidth of concern, the Bragg reflectors are broadband, and both the reflectors and resonator refractive index are non-dispersive.  At the optical carrier frequency $\nu$, the complex reflection response of a lossless FFP formed by two matched reflectors separated by distance $L$, both with amplitude reflection coefficient $r$, can be expressed as
\begin{eqnarray}
  \tilde{F}(\nu) = \tilde{E}_{r}/\tilde{E}_{i} 
         &=& \frac{r(1-exp(-i\theta(\nu)))}{1-r^{2}exp(-i\theta(\nu))} \nonumber \\
         &=& A(\nu)\exp[i\phi(\nu)] ,
\end{eqnarray}
where $\tilde{E}_{r}$ and $\tilde{E}_{i}$ are the reflected and incident electric fields; $\theta(\nu)=2 \pi \nu nL/c$ is the round-trip phase in a material of refractive index n; $A(\nu)$ and $\phi(\nu)$ are, respectively, the amplitude and phase response.  The FFP has a full-width half-maximum (FWHM) bandwidth of $\Delta\nu_{1/2}$.

The PDH locking scheme involves interrogating the FFP with the laser carrier phase modulated at $\nu_{m}$, while measuring the reflected power with a photodetector, as illustrated in Figure \ref{schematic}.  After electronic demodulation and low-pass filtering, this signal can be reduced to \cite{Black}
\begin{eqnarray}
  V(\nu) &\propto& 2\sqrt{P_{c}P_{s}}\times \nonumber \\
               & & \{\Re[\tilde{F}(\nu)\tilde{F}^{*}(\nu_{+})
                   -\tilde{F}^{*}(\nu)\tilde{F}(\nu_{-})]\cos(\psi) \nonumber \\
               & & +\Im[\tilde{F}(\nu)\tilde{F}^{*}(\nu_{+})
                   -\tilde{F}^{*}(\nu)\tilde{F}(\nu_{-})]\sin(\psi)\},
  \label{err_v}
\end{eqnarray}
where the cross term 
\begin{eqnarray}  
  \tilde{C}(\nu_{\pm}) & = & \tilde{F}(\nu)\tilde{F}^{*}(\nu_{+})
                    -\tilde{F}^{*}(\nu)\tilde{F}(\nu_{-})  \nonumber \\
             & = & A(\nu)A(\nu_{+})\exp\{i[\phi(\nu)-\phi(\nu_{+})]\} \nonumber \\
             & & -A(\nu)A(\nu_{-})\exp\{i[\phi(\nu_{-})-\phi(\nu)]\};
  \label{cross_term}                
\end{eqnarray}
$\nu_{+}=\nu+\nu_{m}$ and $\nu{-}=\nu-\nu_{m}$; $P_{c}$ is the power in the carrier while $P_{s}$ is the power in each sideband.  The phase shift $\psi$ is set to optimize the demodulated error signal.  In general this is achieved when
\begin{eqnarray}
  \psi = \tan^{-1}\Bigg\{\frac{d\big[\Im[\tilde{C}(\nu_{\pm})]\big]/d\nu}
         {d\big[\Re[\tilde{C}(\nu_{\pm})]\big]/d\nu}\Bigg\}_{\theta(\nu)=m2\pi},
\end{eqnarray}
where m is an integer.  The round-trip phase $\theta(\nu)=m2\pi$ when the carrier is resonant with the FFP.  

\begin{figure}[htb]
\centering
\includegraphics[width=3.0in]{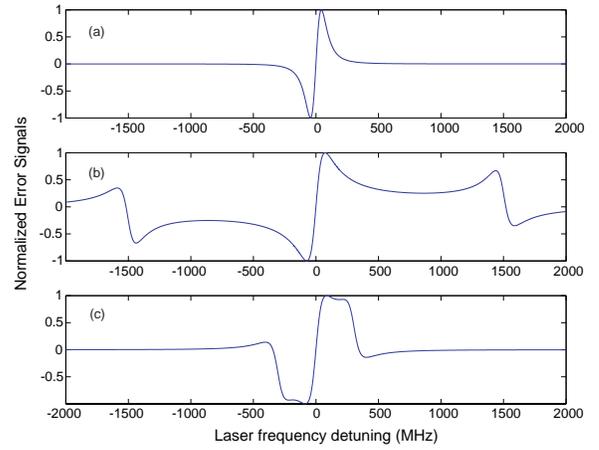}
\caption{Theoretical plots for normalized PDH error signals when an FFP of 150MHz linewidth is interrogated with phase modulation sidebands of (a) 15MHz, (b) 1500MHz, and (c) 300MHz.}
\label{theo_errsig}
\end{figure}

From equation \ref{cross_term}, we can deduce that in the case of $\nu_m \ll \Delta\nu_{1/2}$, $\phi(\nu)-\phi(\nu_{+})$ and $\phi(\nu_{-})-\phi(\nu)$ are both very small, and so the expression is dominated by its real part.  Conversely, when $\nu_m \gg \Delta\nu_{1/2}$, the sidebands are well outside of the FFP linewidth when the carrier is near resonance.  In this case these phase difference terms approach $\pi/2$ and the expression is dominated by its imaginary part.  If the FFP lineshape is symmetric and the carrier is at resonance, $A(\nu_{+})=A(\nu_{-})$ and $\phi(\nu)-\phi(\nu_{+})=\phi(\nu_{-})-\phi(\nu)$ for both cases, implying that equation \ref{cross_term}, and hence equation \ref{err_v}, become zero.  This is the usual lock point of the frequency servo.  From equation \ref{err_v}, it is clear that when the cross term equals 0 (locked to resonance), the output $V(\nu)$ is equal to zero and independent of $P_{c}$ and $P_{s}$.  Hence, when locked, the PDH system is immune to variations in laser intensity noise to the first order.  In comparison, a fringe-side locking technique shows no implicit immunity to intensity noise, and requires an additional intensity monitor and subtraction electronics \cite{Lissak}.

Figure \ref{theo_errsig}a illustrates the theoretical error signal for the case of $\nu_{m}/\Delta\nu_{1/2}=0.1$, while Figure \ref{theo_errsig}b is for the case of $\nu_m/\Delta\nu_{1/2}=100$, when $\nu$ is scanned across the resonance of a FFP.  Figure \ref{theo_errsig}c shows the intermediate case where $\nu_m/\Delta\nu_{1/2}=2$.  The two satellite error signals in Figure \ref{theo_errsig}b are due to the sidebands undergoing the FFP resonance, whereas in Figure \ref{theo_errsig}c, the error signals due to the carrier and sidebands merge to form a single and almost square error signal.  The plots assume a resonance linewidth of 150MHz, and it is interrogated by phase modulation frequencies 15MHz, 1500MHz and 300MHz respectively.

\begin{figure}[htb]
\centering
\includegraphics[width=3.0in]{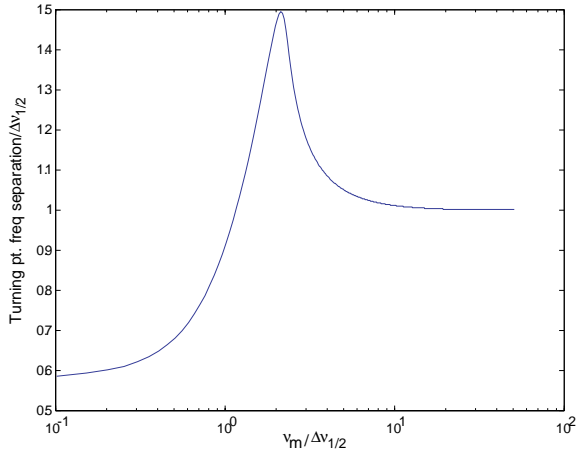}
\caption{Theoretical plot for the frequency seperation of the error signal turning points vs. modulation frequency.  Both axes are normalized by $\Delta\nu_{1/2}$.}
\label{theo_errsig_bw}
\end{figure}

The case where $\nu_m \gg \Delta\nu_{1/2}$ describes the classic PDH locking regime, involving high finesse Fabry-Perot cavities.  The principle of operation behind both extremes are similar and, for the sake of brevity, we shall refer to both as PDH locking in this treatment.  Subsequent experimental results for our FFP to be presented in this paper will show that we were operating nearer the $\nu_m \ll \Delta\nu_{1/2}$ regime.  

\begin{figure}[htb]
\centering
\includegraphics[width=3.0in]{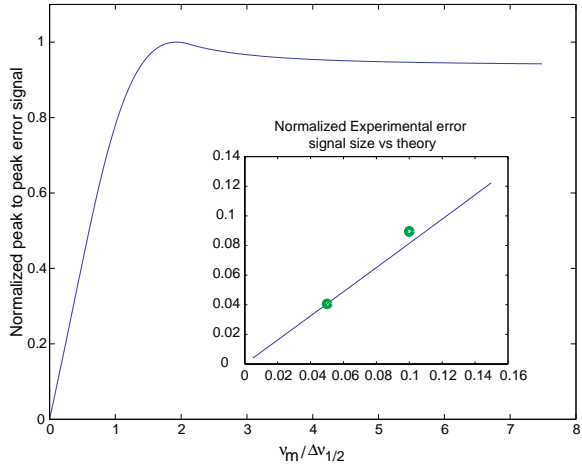}
\caption{Theoretical plot for the normalized peak-to-peak error signal vs. modulation frequency, normalized by $\Delta\nu_{1/2}$.  Inset: Normalized experimental operating regimes for two resonances, overlaid with expanded theoretical plot.}
\label{theo_errsig_size}
\end{figure}

For a given resonance FWHM, $\Delta\nu_{1/2}$, the frequency separation between the turning points of a PDH error signal is dependent on $\nu_{m}$.  It approaches asymptotic values for both cases of $\nu_m \ll \Delta\nu_{1/2}$ and $\nu_m \gg \Delta\nu_{1/2}$, as illustrated by the theoretical plot in Figure \ref{theo_errsig_bw}.  The plot is calculated with $\psi$ optimized for each $\nu_{m}$.

On the other hand, for a given modulation frequency, the size and, therefore, slope of the error signal is dependent on the FWHM bandwidth $\Delta\nu_{1/2}$.  Figure \ref{theo_errsig_size} shows the theoretical plot of peak-to-peak normalized error signal size vs normalized FWHM bandwidth.  The error signal size approaches zero when $\nu_{m}\ll\Delta\nu_{1/2}$, but reaches an asymptotic value when $\nu_{m}\gg\Delta\nu_{1/2}$.

\section{Experimental setup}
The topology of our PDH interrogation of an FFP is shown in Figure \ref{schematic}.  The laser carrier is provided by a New Focus Vortex 6029, which was an extra-cavity diode laser with a factory-estimated linewidth of 1MHz, and an intrinsic linewidth of $\approx$300kHz.  Its optical wavelength was centered around 1550.15nm, with about 0.40nm tuning range, which corresponds to a frequency range of $\approx$50GHz.  The frequency tuning of the laser was actuated by applying a voltage to the piezo-electric transducer (PZT), which changed the laser cavity length.  The factory calibration specified that its laser PZT actuator had a gain of 12.5GHz/V.

\begin{figure}[htb]
\centering
\includegraphics[width=3.0in]{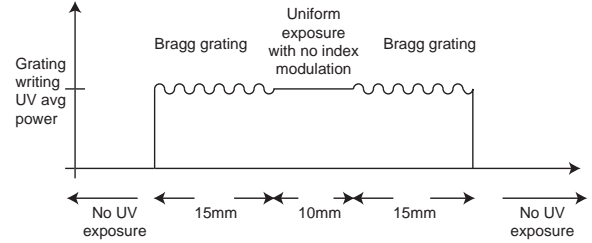}
\caption{Schematic for the UV exposure, with no apodization, along the length of a fiber Fabry-Perot.}
\label{FFP_schematic}
\end{figure}

After passing through the optical isolator, the laser polarization was adjusted to vertical by a half-wave plate before being modulated at 15MHz by the resonant phase modulator (New Focus 4003).  The phase modulator was driven by a radio-frequency (RF) signal generator, which also provided the local oscillator for the demodulation electronics.  The modulated laser beam was coupled with an aspheric lens into a fiber-pigtailed polarization-independent optical circulator, which was spliced to the FFP.  The FFP was held between a pair of magnetic clamps, with one of the clamps in turn mounted on a translation stage, so that the Bragg wavelength could be stretch-tuned to within the laser frequency range.  Our FFP consisted of a pair of nominally matched 13.5dB Bragg gratings (R$\approx95.5\%$) each 15mm long, spaced 10mm apart, fabricated in a single phase-coherent writing process.  The schematic for the UV exposure along the length of the fiber is illustrated in Figure \ref{FFP_schematic}.  They were written in hydrogenated SMF-28 fiber with no apodization.  Both the transmitted and reflected light were collimated back into free space with ashperic lenses and then focussed onto photodetectors Tx and Rx, respectively, each with electronic bandwidth of $\approx$20MHz.  The optical isolator in the transmitted port eliminated any parasitic etalon effects due to residual back reflections from the collimating asphere.  The RF local oscillator was phase shifted before being used to mix down the electronic signal from the reflected port.  The mixed signal was low-pass filtered to provide the PDH error signal.  The local oscillator phase shift $\psi$ was optimized experimentally by maximizing the error signal.

\section{Experimental results and discussion}

\subsection{High resolution fiber Fabry-Perot characterization by laser frequency scanning}
\begin{figure}[htb]
\centering
\includegraphics[width=3.0in]{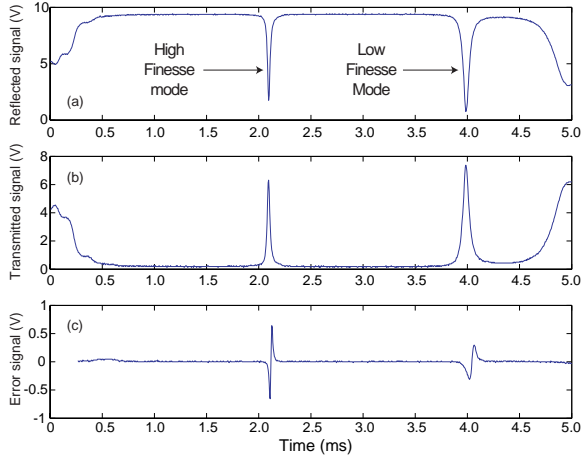}
\caption{Experimental scans for (a) reflection, (b) transmission, and (c) PDH error signal for our fiber Fabry-Perot.}
\label{exp_scan_wide}
\end{figure}

A 95Hz voltage ramp of 2Vp-p and 50:50 symmetry was applied to the laser PZT input to sweep the laser carrier frequency, which equates to a slope of 380V/s.  The intensities transmitted and reflected by the FFP, as measured by the photodetectors, and the corresponding mixed down experimental error signal were recorded using a digital oscilloscope while the laser frequency was scanned.  They are displayed in Figures \ref{exp_scan_wide}a, \ref{exp_scan_wide}b and \ref{exp_scan_wide}c, respectively.  There were two FFP resonances within the Bragg grating bandwidth with differing peak heights and $\Delta\nu_{1/2}$'s.  These differences were mainly due to the frequency dependent reflectivity of the Bragg grating pair, thus resulting in differing finesse at the two resonances.

Since the gratings were not apodized during the fabrication process, we can expect higher reflectivity near the center of their bandwidth, which is confirmed by the higher finesse and thus narrower FWHM of the first resonator mode.  Further, by comparing the heights of the two peaks in Figure \ref{exp_scan_wide}a, we can see that the lower finesse resonance is closer to impedance matching.  At this mode, nearly all of the laser light was transmitted and the reflection approached zero.  This difference in transmitted intensity, compared with the under-coupled high finesse mode, can be explained by UV induced loss in the resonator, particularly in the 10mm spacing between the grating pair.  The higher finesse resonance transmitted less intensity due to its greater resonator round-trip number, or total storage time, which resulted in greater total loss while circulating within the resonator.  To reduce this loss, the UV laser can be easily controlled to avoid fiber exposure between the grating pair during the resonator fabrication process.  

\begin{figure}[htb]
\centering
\includegraphics[width=3.0in]{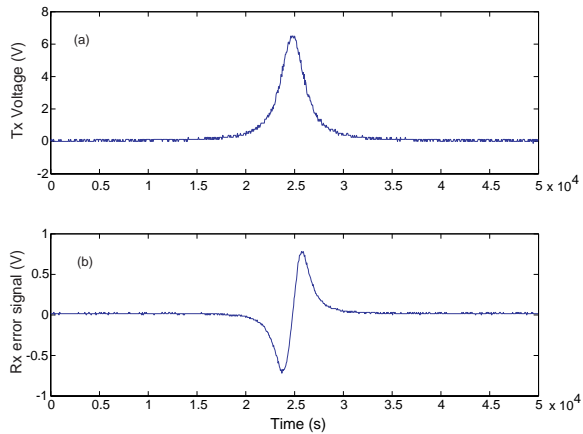}
\caption{Experimental scans for (a) transmission, and (b) reflected PDH error signal for our fiber Fabry-Perot.}
\label{exp_scan_zoom}
\end{figure}

The transmission scan and the reflected error signal for the narrower resonance is enlarged in Figure \ref{exp_scan_zoom}a and \ref{exp_scan_zoom}b.  The FWHM time for the PZT scan in Figure \ref{exp_scan_zoom}a was $\simeq30\mu$s, which corresponds to 11.4mV on the PZT.  Recalling that the factory calibration specified that its laser PZT input provided 12.5GHz/V of tuning, the FWHM bandwidth of this mode can be determined to be 143MHz.  For comparison, the broader resonance had a FWHM time of 66$\mu$s, which implies a $\Delta\nu_{1/2}$ bandwidth of 314MHz.  The separation between the two peaks can be seen to be $\simeq1.9$ms in Figure \ref{exp_scan_wide}, which infers a free spectral range of 9GHz.  Hence, the narrower mode had a finesse of 63, while the broader resonance had a finesse of 29.

The $\nu_{m}/\Delta\nu_{1/2}$ ratio for the higher finesse mode was $\simeq0.1$.  The corresponding peak-to-peak time for its error signal in Figure \ref{exp_scan_zoom}b was $\simeq 20\mu$s, which yields an error signal turning point frequency separation to $\Delta\nu_{1/2}$ ratio of $\simeq0.60$.  On the other hand, the lower finesse resonance had an error signal peak-to-peak time of 38$\mu$s, which corresponds to $\nu_{m}/\Delta\nu_{1/2}$ of $\simeq0.05$, and an error signal turning point separation to $\Delta\nu_{1/2}$ ratio of $\simeq0.58$.  The error signal turning point separation to $\Delta\nu_{1/2}$ ratios for the two modes are close to each other, and agree with the values as predicted in Figure \ref{theo_errsig_bw}.  At these linewidths, $\nu_{m}$ is small enough relative to $\Delta\nu_{1/2}$ to approach the asymptotic value of the lower limit.

The peak-to-peak error signal size for the higher finesse mode was larger than that of the lower one, as seen in Figure \ref{exp_scan_wide}c, since the $\nu_{m}/\Delta\nu_{1/2}$ for the higher finesse mode was twice that of the lower finesse mode.   This was predicted by the theoretical plot in Figure \ref{theo_errsig_size}.  The error signal peak-to-peak voltage for the high finesse mode was measured to be 1.4V, while that for the lower finesse resonance was 0.63V.  These two points, for $\nu_{m}/\Delta\nu_{1/2}$ of 0.1 and 0.05, are normalized and overlaid with the theoretical plot in the inset of Figure \ref{theo_errsig_size}, to illustrate the region where these two modes were operated.

Assuming an effective refractive index of 1.45, a free spectral range of 9GHz yields a resonator length of 11.5mm, implying that the effective reflection point of the gratings was $\simeq 0.75$mm inside each grating.

We tested for polarization dependence of the FFP response with a second half-wave plate before the laser was coupled into the fiber.  No visible shift in resonance frequencies were observed as the waveplate was rotated.  This implies that for the intent and purpose of this application, the UV illumination of the grating core during the fabrication process can be regarded as isotropic.  Any degeneracy due to parasitic birefringence was beyond the linewidth resolution of the FFP resonance, as the two modes provided well behaved error signals free from input polarization wander effects.

It is evident from Figures \ref{exp_scan_wide} and \ref{exp_scan_zoom} that PZT scanning to sweep the frequency of the laser, as demonstrated in this experiment, is a simple alternative to the single-sideband modulation technique for high resolution spectral characterization of fiber gratings \cite{Roman}.

The slope of the error signal through resonance was $\simeq19$nV/Hz for the higher finesse mode, and $\simeq9$nV/Hz for the lower finesse mode.  Hence the higher finesse resonance was our preferred mode for PDH locking, as it provided a larger signal as a frequency and displacement discriminator in sensing applications.  One should note, however, that while higher FFP finesse is preferred for superior sensitivity, the free running laser frequency noise sets a limit to interferometer sensitivity.

\subsection{Sensor signal extraction in frequency locked operation}
\begin{figure}[htb]
\centering
\includegraphics[width=3.0in]{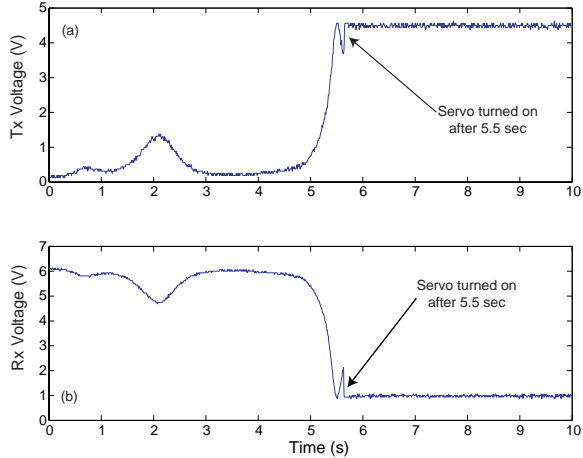}
\caption{Oscilloscope traces for (a) transmitted, and (b) reflected intensities during lock acquisition.  The feedback loop was engaged at $\approx$5.5sec.}
\label{exp_lock_acq}
\end{figure}

To lock the laser, the voltage ramp from the signal generator was turned off, and the PZT DC offset voltage tuned slowly while the transmitted and reflected laser intensities were monitored with an oscilloscope.  When the laser is nearly resonant with the chosen FFP peak, the transmitted intensity approaches its maximum, and the feedback loop was then engaged to acquire lock.  This process was recorded by the digital oscilloscope traces shown in Figure \ref{exp_lock_acq}.  The servo amplifier used in this experiment had a single real pole response with a corner frequency of 0.03Hz.  The total feedback loop had a DC gain of $\approx$1000 and a unity gain bandwidth of around 40Hz.  Lock acquisition was straight forward and once it was acquired, the system stayed locked for several hours even when subjected to large environmental noise events.  Lock termination occured when the grating drifted outside the laser tuning range, and this typically happened after over 3 hours of locked operation.

\begin{figure}[htb]
\centering
\includegraphics[width=3.0in]{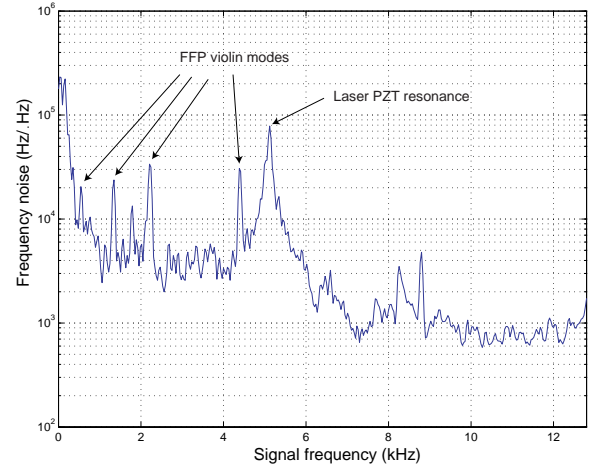}
\caption{Low frequency noise of a fiber Fabry-Perot sensor as measured by a dynamic signal analyzer.}
\label{exp_sig_analyser_lowfreq}
\end{figure}

In a PDH locking scheme, the sensor signal is extracted by either monitoring the PZT feedback voltage required to maintain lock within the servo unity gain bandwidth, or by monitoring the mixer output at frequencies above the unity gain bandwidth.  Environmental stimulations, such as temperature drift as well as stress and strain due to mechanical or acoustic perturbation, change the resonance condition of the FFP.  This results in both DC and AC voltage change in the mixer output and the PZT feedback voltage.  When the mixer output is read with a dynamic signal analyzer, information about these perturbations can be extracted.  The signal analyzer performs a Fast Fourier Transform of the mixer output voltage, and provides a trace with units in Volts$/\sqrt{\textrm{Hz}}$.  The quotient of this trace by the slope of the error signal (19nV/Hz) yields the callibrated measurement in Hz$/\sqrt{\textrm{Hz}}$.  The low frequency measurement of this mixer output is shown in Figure \ref{exp_sig_analyser_lowfreq}.  There was a large component of ambient noise at low frequencies as the FFP was not isolated from laboratory acoustic and thermal noise.  We were able to identify fiber violin modes, broadband acoustic noise, and PZT resonances in this frequency regime.  For example, the large feature at $\sim 5$ kHz seen in Figure \ref{exp_sig_analyser_lowfreq} is due to closed loop excitation of a laser PZT mode.

Figure \ref{exp_sig_analyser} shows a wider frequency scan of the ambient frequency noise. It is overlaid with the calculated shot noise and measured electronic noise.  At frequencies above ambient excitation, the free running frequency noise of the laser limits this measurement to $\simeq$300Hz$/\sqrt{\textrm{Hz}}$.  Assuming the laser has a Lorentzian lineshape with white spectral density of frequency noise $S_{f}$, the 3dB linewidth of the laser $\Delta\nu_{L}$ can be estimated by \cite{Lissak,Elliott,Zhang}
\begin{eqnarray}
   \Delta\nu_{L} & = & \pi S_{f}^{2},
\end{eqnarray}
where $S_{f}$ has units of Hz$/\sqrt{\textrm{Hz}}$.  Thus, the broadband frequency noise of $\simeq$300Hz$/\sqrt{\textrm{Hz}}$ corresponds to an intrinsic laser linewidth of $\simeq280$kHz, which is consistent with the manufacturer's estimate of 300kHz.  

\subsection{Strain sensitivity, and dynamic range}
According to the empirical model determined by Kersey et al. \cite{FBGsensorsKersey}, Bragg grating responsivity
\begin{eqnarray}
    \frac{1}{\lambda_{B}}\frac{\delta\lambda_{B}}{\delta\varepsilon} & = &
        0.78 \varepsilon^{-1},
    \label{Kerseymodel}
\end{eqnarray}
where $\varepsilon$ is the strain perturbation, and $\lambda_{B}$ is the Bragg wavelength, 1pm of induced grating wavelength shift corresponds to a strain of $\simeq 0.8 \mu\varepsilon$.  At $\lambda_{B}=1550$nm, equation (\ref{Kerseymodel}) can be rearranged to arrive at the conversion factor
\begin{eqnarray}
    \frac{\delta\varepsilon}{\delta\nu_{B}} & = & \frac{\lambda_{B}}{0.78c} \nonumber \\
      & = & 6.6 \times 10^{-15} \varepsilon/\textrm{Hz},
\end{eqnarray}
where $\delta\nu_{B}$ is the equivalent induced grating frequency shift.  Since 1pm is equivalent to 125MHz at 1550nm, we can infer from the high frequency noise floor that the FFP sensor has a broadband strain sensitivity of $\approx 2 p\varepsilon/\sqrt{\textrm{Hz}}$.

\begin{figure}[htb]
\centering
\includegraphics[width=3.0in]{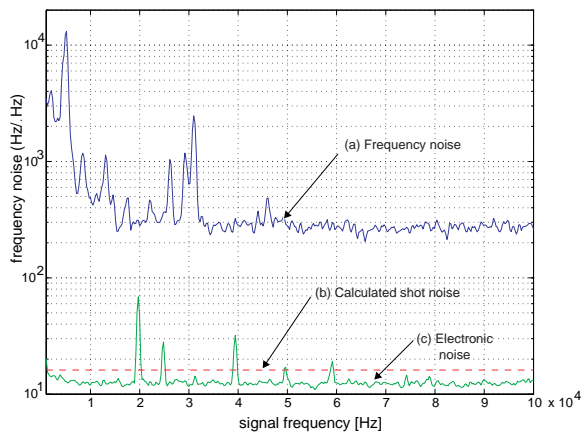}
\caption{(a) Frequency noise of a fiber Fabry-Perot sensor as measured by a dynamic signal analyzer, overlaid with (b) calculated shot noise, and (c) measured electronic noise.}
\label{exp_sig_analyser}
\end{figure}

The shot noise in Figure \ref{exp_sig_analyser} was calculate as follows \cite{Yariv}:
\begin{eqnarray}
    V_{SN} & = & \alpha\sqrt{2egV_{dc}} \qquad \textrm{V}_{\textrm{rms}}/\sqrt{\textrm{Hz}},
\end{eqnarray}
where $V_{SN}$ is the equivalent shot noise voltage; $e=1.602\times10^{-19}$ C is the electronic charge; $V_{DC}$ is the $dc$ output voltage of the photodetector when the system is locked; g is the transimpedance gain of the photodetector: and $\alpha$ is the mixer conversion gain.  The quotient of $V_{SN}$ by the error signal slope then gives the shot noise in units of Hz$/\sqrt{\textrm{Hz}}$.  This was calculated to be 16 Hz$/\sqrt{\textrm{Hz}}$, which corresponds to a limiting shot-noise sensitivity of (16 Hz$/\sqrt{\textrm{Hz}} \times \delta\varepsilon/\delta\nu_{B}$) $\simeq$ 100 f$\varepsilon/\sqrt{\textrm{Hz}}$.  The electronic noise is the dark noise measured at the mixer output.

Within the unity gain bandwidth of the feedback system, the sensor dynamic range depends on the laser optical frequency tuning range.  Since our laser had a PZT tuning range of 50 GHz, the low frequency dynamic range of this system is limited to ($50 \times 10^{9} \textrm{Hz} \times \delta\varepsilon/\delta\nu_{B} = $) 330 $\mu\varepsilon$.  Assuming a breaking stress of $>$ 100 kpsi \cite{Carter}, and a Young's modulus of $1.02 \times 10^{4}$ kpsi \cite{Corning} for fused silica, the breaking strain is $>$ 9800 $\mu\varepsilon$.  This means that typically, the breaking strain is well beyond the limited tuning range of the laser used in this experiment.  Above the unity gain bandwidth, the sensor dynamic range is limited by the FWHM bandwidth of the resonator to ($143 \times 10^{6} \textrm{Hz} \times \delta\varepsilon/\delta\nu_{B} = $) 0.9 $\mu\varepsilon$.  Hence, for large dynamic range applications, the preferred operating approach is to expand the unity gain bandwidth out to a maximum, and perform in-loop measurements at the laser PZT actuator input.

\section{Conclusion}
We have presented a passive fiber sensor interrogation technique which was adapted from PDH locking, used in gravitational wave detection.  We demonstrated the robust and stable operation of the simple locking system in fiber.  In many applications, we believe this passive technique is superior to active methods using fiber lasers, due to its better efficiency, improved signal to shot-noise ratio, lower cost, and its suitability for remote sensing.  It has an implied broadband strain sensitivity limit of 2 p$\varepsilon$, which is due to the free-running frequency noise of our laser.  With appropriate laser stabilization prior to FFP interrogation, however, it has the potential to surpass the pico-strain regime and approach the fundamental shot noise limit.

 
\hfill June 29, 2004

\section*{Acknowledgment}
The authors would like to thank Adrian L. G. Carter, of Nufern, for useful discussions.

\end{document}